\documentclass[conference]{IEEEtran}

\IEEEoverridecommandlockouts

\makeatletter
\let\old@ps@headings\ps@headings
\let\old@ps@IEEEtitlepagestyle\ps@IEEEtitlepagestyle
\def\confheader#1{%
\def\ps@headings{%
\old@ps@headings%
\def\@oddhead{\strut\hfill#1\hfill\strut}%
\def\@evenhead{\strut\hfill#1\hfill\strut}%
}%
\def\ps@IEEEtitlepagestyle{%
\old@ps@IEEEtitlepagestyle%
\def\@oddhead{\strut\hfill#1\hfill\strut}%
\def\@evenhead{\strut\hfill#1\hfill\strut}%
}%
\ps@headings%
}
\makeatother

\usepackage{cite}
\usepackage{amsmath,amssymb,amsfonts}
\usepackage{algorithmic}
\usepackage{graphicx}
\usepackage{textcomp}
\usepackage{xcolor}
\def\BibTeX{{\rm B\kern-.05em{\sc i\kern-.025em b}\kern-.08em
    T\kern-.1667em\lower.7ex\hbox{E}\kern-.125emX}}
\begin{document}
\bstctlcite{IEEEexample:BSTcontrol}

%Title
\title{Smart Grid Cyber Attacks Detection using Supervised Learning and Heuristic Feature Selection}

\confheader{%
To appear in the proceeding of IEEE SEGE 2019.
}

\author{\IEEEauthorblockN{Jacob Sakhnini}
\IEEEauthorblockA{\textit{School of Engineering} \\
\textit{University of Guelph}\\
Guelph, Ontario, Canada \\
jsakhnin@uoguelph.ca}
\and
\IEEEauthorblockN{Hadis Karimipour}
\IEEEauthorblockA{\textit{School of Engineering} \\
\textit{University of Guelph}\\
Guelph, Ontario, Canada \\
hkarimi@uoguelph.ca}
\and
\IEEEauthorblockN{Ali Dehghantanha}
\IEEEauthorblockA{\textit{School of Computer Science} \\
\textit{University of Guelph}\\
Guelph, Ontario, Canada \\
adehghan@uoguelph.ca}
}

\maketitle

%Abstract
%-----------------------------------------
\begin{abstract}
False Data Injection (FDI) attacks are a common form of Cyber-attack targetting smart grids. Detection of stealthy FDI attacks is impossible by the current bad data detection systems. Machine learning is one of the alternative methods proposed to detect FDI attacks. This paper analyzes three various supervised learning techniques, each to be used with three different feature selection (FS) techniques. These methods are tested on the IEEE 14-bus, 57-bus, and 118-bus systems for evaluation of versatility. Accuracy of the classification is used as the main evaluation method for each detection technique. Simulation study clarify the supervised learning combined with heuristic FS methods result in an improved performance of the classification algorithms for FDI attack detection.
\end{abstract}


%Introduction
%------------------------------------------------------
\section{Introduction}
\label{sec:intro}
Today's power systems consist of a network of sensors and generators that allow two way communication within the system's infrastructure. This feature allows utility companies to distribute power more efficiently along larger areas by real time demand side management. While this complex communication system has tremendous advantage, it is more prone to measurement tampering and cyber-attacks. These cyber-attacks come in various forms and are typically constructed for the purpose of power theft or causing power outages and disturbances. False Data Injection (FDI) is a form of cyber-attack in which the measurements are altered in a stealthy manner \cite{100}. Such attacks can bypass the standard defence mechanisms used today; and as such, machine learning, among other methods, are proposed to detect these attacks.

FDI attacks are common form of cyber-attacks targeting smart grids. The danger in these types of attacks stems from their ability to bypass the standard state estimation system used in most smart grids \cite{200}. The inability to detect these attacks through state estimation creates the need for other methods for classifying these attacks. FDI detection is typically achieved through analysis of meter measurements throughout the power system. Various detection methods have been proposed that rely on spatial-temporal correlation, real-time correlation, and statistical correlation of meter measurements. \cite{300}\cite{FDImethods} studies the three methods and suggests that detection of FDI attacks based on real-time correlation is favourable to intelligent machine learning techniques due to its ability to scale to larger systems with low computational cost. However, appropriate feature and parameter selection can greatly improve the computational efficiency of any machine learning algorithm.

Several machine learning based approaches for detecting FDI attacks have been proposed in literature \cite{SVManomaly}\cite{400}. \cite{SVManomaly} compared a supervised and a non-supervised approach by using support vector machines and anomaly detection algorithms. It concluded that both machine learning algorithms are successful at detecting FDI attacks based on statistical deviations in measurements. Furthermore, it concluded that the features used in detecting these attacks are highly correlated and can be reduced to two dimensions with Principal Component Analysis (PCA) while retaining 0.99 of the variance. While correlation is expected in grid measurements, such high correlation was found in this study due to the low variation in the simulation of data. With more thorough simulation, the complexity of the problem increases and correlation is expected to decrease. Furthermore, larger power systems are expected to have less co-variance among the measurements; and as such, alternative feature selection (FS) methods are necessary.

\cite{OzayML} has tested and compared more algorithms for detection of FDI attacks. The supervised learning algorithms used in this study are linear and Gaussian SVM, K Nearest Neighbour (KNN), and a single-layer perceptron. The study concluded that KNN is more sensitive to the system size and may perform better in small size systems. It also concluded that SVM performs better on large-scale systems, specifically with a Gaussian kernel. Single layer perceptron was also observed to be less sensitive to the system size, however not as accurate as SVM. A multi-layered perceptron, also known as an artificial neural network (ANN), is hypothesized to be more accurate due to its increased complexity.

Similarly, \cite{SVMandKNN} tested SVM, KNN, and Extended Nearest Neighbours (ENN), and compared their accuracy on the IEEE 30-bus system. General conclusions can be made about the success of machine learning in classifying FDI attacks. However, this study lacks thorough cross-validation between algorithms of varying parameters. Furthermore, testing was only done on one system, so no conclusions can be made on the versatility of the classification algorithms among power systems of varying sizes.

In this paper, the performance of three different classification techniques are tested with three heuristic FS techniques. The three machine learning algorithms used are SVM, KNN algorithm, and ANN. The three FS techniques are Binary Cuckoo Search (BCS), Binary Particle Swarm Optimization (BPSO), and Genetic Algorithm (GA). The goal is to combine machine learning and FS techniques to take advantages of their strength and compensate their weaknesses. These algorithms will be compared based on their classification accuracy and computational efficiency. The results show that heuristic FS techniques are capable of selecting a subset of features that can obtain a higher classification accuracy with a significantly lower number of features.

%System Model
%----------------------------------------------------
\section{System Model}
\label{sec:SysModel}
\subsection{State Estimation of Power Systems}
\label{ssec:sys_StateEstimation}
Power systems that employ smart grid technologies rely on state estimation to predict the state of the system which determines the optimal power generation. This technique represents a relationship between the state variables of the system and the real measurements recorded along the power grid \cite{StateEstimation}\cite{500}. The measurement data consists of power flow, voltage magnitude and phase angles described as follows:

\begin{equation}
Z(k)=H(k) x(k) + \epsilon(k)
\end{equation}

where $Z$ represents measurement vector, $x$ represents vector of state variables, $H$ is the Jacobian matrix, and $\epsilon$ is the measurement error. $k$ refers to the time step.
The state estimation problem under the assumption of global observability can be formulated using the least squares method as follows:

\begin{equation}\label{e:eq3}
\begin{aligned}
  \hat {{x}}(k+1)=\hat {{x}}(k)+{G}^{-1}(k){H}(k){W}^{-1} [{{Z}}(k)-{H}(k)\hat {{x}}(k)]  ,\\
\end{aligned}
\end{equation}
where gain matrix ${G}(k)={H}^{T}(k){W}^{-1} {H}(k)$. $\hat {{x}}$ is the vector of estimated states of the system. ${W}$ is the covariance matrix.
To make sure about the accuracy of the estimation, measurement data will be checked to remove bad data \cite{600}.Traditionally, bad data is detected through following 2-norm residual test:

\begin{equation}
   \|z-H x\|^{2}<\varepsilon 
\end{equation}
where $\varepsilon$ is the threshold for bad data detection. If the residual of the measurements go above the predefined threshold bad data exist and should be removed before the next iteration.

\subsection{False Data Injection Attacks}
\label{ssec:sys_FDI}
FDI attacks consist of malicious data injected into the measurement meters. FDI attacks can be performed by manipulating the measurements along the network by a linear factor of the Jacobian matrix of the system \cite{FDImethods} \cite{FDI}. An FDI attack can then be simulated as

\begin{equation}
    Z_{b a d}=Z+a
\end{equation}
where a is an attack vector such that $a = Hc$ which results in
$$
\|Z-H x\|^{2}=\left\|Z_{b a d}-H x_{b a d}\right\|^{2}+\Gamma
$$
where $\Gamma$ is an error term attributed to the state estimation that must remain within a certain threshold depending on the power system.

%Supervised Learning Based Detection of FDI
%----------------------------------------------------
\section{Supervised Learning Based Detection of False Data Injection}
\label{sec:SupervisedLearning}
Three supervised classification algorithms will be cross validated along with three heuristic FS techniques. The following sections discuss the algorithms implemented in this paper.

\subsection{Feature Selection}
\label{ssec:SL_featSelect}
Power systems are highly complex and large scale physical systems with huge number of  feature and measurements. Therefore, feature selection is an essential task that should be performed to optimize the computational efficiency \cite{700}. Principal Component Analysis (PCA) has been used in previous literature for dimensionality reduction \cite{SVManomaly}. However, large-scale power systems behave somewhat non-linearly; and as such, heuristic approaches to feature selection are considered. In this paper, GA, Cuckoo Search (CS), and Particle Swarm Optimization (PSO) are used to increase the computational efficiency of the supervised learning algorithms. Each of the algorithms are aimed to obtain the most optimal subset of features that results in the best accuracy. Each solution consists of a binary vector with each index being 1 if the feature is used in this subset and 0 if it is not.

\subsubsection{Binary Cuckoo Search}
\label{sssec:BCS}
BCS is a binary implementation of CS, an optimization algorithm based on the parasite behavior of some species of Cuckoo. The CS algorithm is proposed by \cite{CSoriginal} and summarized by the following three rules:

\begin{enumerate}
    \item Each Cuckoo lays one egg at a randomly chosen nest.
    \item The best nests with high quality eggs carry over to the next generation.
    \item The number of available nests is fixed. And if another cuckoo egg is discovered by the host bird, the host can remove the egg or build a new nest.
\end{enumerate}

Mathematically, the nests, or solutions, are updated using random walk via Lévy flights:

\begin{equation}
x_{i}^{j}(t)=x_{i}^{j}(t-1)+\alpha \oplus L e v y(\lambda)
\end{equation}
and
\begin{equation}
L e v y \sim u=s^{-\lambda},(1<\lambda \leq 3)
\end{equation}
where $x_{i}^{j}$ is the $j^{th}$ egg (feature) at nest (solution) $i$, $s$ is the step size, $\alpha > 0 $ is the step size scaling factor, and $\oplus$ is the entry-wise product. The Lévy flights employ a random step length which is drawn from a Lévy distribution which creates longer step length in the long run allowing more efficient search space exploration \cite{CSoriginal}. The solutions are restricted to binary values by the following equations:

\begin{equation}
S\left(x_{i}^{j}(t)\right)=\frac{1}{1+e^{-x_{i}^{j}(t)}}
\end{equation}
\begin{equation}
x_{i}^{j}(t+1)=\left\{\begin{array}{ll}{1} & {\text { if } S\left(x_{i}^{j}(t)\right)>\sigma} \\ {0} & {\text { otherwise }}\end{array}\right.
\end{equation}
in which $\sigma \sim U(0,1)$ and $x_{i}^{j}(t)$ denotes the new egg value at time $t$ \cite{BCS}.

\subsubsection{Genetic Algorithm}
\label{sssec:GA}
GA is an optimization technique that yields the best solution based on the evolution mechanism of living beings \cite{GAfdi}. Following the principle of natural selection, GA chooses the best solutions based on their fitness. In each iteration, GA eliminates the solutions with the lowest fitness and retains the solutions with the highest fitness. Similarly to \ref{sssec:BCS}, the solution consists of a binary vector indicating the variables used as features, and the fitness of each solution is the classification accuracy of FDI attacks based on that subset of features.

\subsubsection{Binary Particle Swarm Optimization}
\label{sssec:BPSO}
PSO is an algorithm used for solving a variety of problems. The algorithm is motivated by social behaviours in nature. The main characteristic of this algorithm is that optimization is performed through social interaction in the population where thinking is not only personal, but also social \cite{PSOclassification}. A binary implementation of Particle Swarm Optimization (BPSO) is also used as a heuristic method for feature selection. 

The first step of implementing BPSO is initialization of population consisting of user defined particles; each particle represents a feasible solution. Through iterations, particles update themselves by tracking two criteria. The first criterion is the best solution of each particle. Personal best of the $i^{th}$ particle is $p B e s t_{\mathrm{i}}=\left(p B e s t_{i}^{1}, p B e s t_{i}^{2}, \ldots, p B e s t_{i}^{n}\right)$. And the second criterion is global best solutions, $g B e s t=\left(g B e s t^{1}, g B e s t^{2}, \ldots, g B e s t^{n}\right)$ respectively.

\subsection{Classification Algorithms}
\label{ssec:ClassificationAlgorithms}
Three types of supervised learning algorithms are implemented in this study for the purpose of cross-validation and analysis. The three types of classification algorithms use different mathematical approaches to classify the data. The following subsections will explain each of the algorithms to be implemented.

\subsubsection{Support Vector Machine}
\label{sssec:SVM}
SVM is an algorithm that classifies data by constructing a set of hyper-planes in high dimensions \cite{SVM}. To simplify the computations, kernel functions are used to represent the mapping of the data. In this study, a Gaussian kernel will be used for the SVM due to its non-linear properties and its capability of classifying data based on statistical variances with high computational efficiency. Mathematically, the Gaussian kernel is defined as follows:

\begin{equation}
K\left(x_{i}, x_{i^{\prime}}\right)=\exp \left\{-\gamma \sum_{j=1}^{p}\left(x_{i j}-x_{i^{\prime} j}\right)^{2}\right\}
\end{equation}

where $\gamma$ is the kernel coefficient. The SVM algorithm will be tested with varying penalty parameter, $C$, and kernel coefficient, $\gamma$, and cross-validated for accuracy.

\subsubsection{K- Nearest Neighbours}
\label{sssec:knn}
KNN algorithm classifies data based on its closest $k$ neighbours. The closeness between the data is determined using the euclidean distance,

\begin{equation}
d_{i j}=\left\|\mathbf{s}_{\mathbf{i}}-\mathbf{s}_{j}\right\|, \mathbf{s}_{j} \in S
\end{equation}

where $S$ and $s$ correspond to labelled and unlabelled data respectively. For $k>1$, data is classified based on majority of neighbours. In this study, various k values will be tested and cross validated for accuracy.

\subsubsection{Artificial Neural Network}
\label{sssec:ann}
ANN is an algorithm composed of interconnected elements, called neurons or nodes, which process information based on specific weights. ANNs can be constructed in various methods and architectures and typically consist of an input layer, hidden layers, and an output layer each consisting of several nodes. Each node $i$ performs calculations represented by the transfer function $f_i$ as follows:

\begin{equation}
y_{i}=f_{i}\left(\sum_{j=1}^{n} w_{i j} x_{j}-\theta_{i}\right)
\end{equation}

where $y_i$ is the output of the node $i$, $x_j$ is the $j^{th}$ input to the node, $w_{ij}$ is the connection weight between nodes $i$ and $j$, and $\theta_{i}$ is the bias of node $i$.

The architecture of the ANNs implemented in this study consist of an input layer of $L$ nodes, one hidden layer of $M$ nodes, and an output layer of $N$ nodes; where $L$ is equal to the number of features in the input data, $N$ is equal to 2, the number of classes, and $M$ is calculated as follows:

\begin{equation}
M=\left\lceil\frac{N+L}{2}\right\rceil
\end{equation}

The ANN algorithm will be implemented with varying learning rate, $\alpha$, and using back propagation as a learning solver.

\subsection{Evaluation Methods}
\label{ssec:meth_eval}
The classification algorithms implemented in this study are evaluated based on their prediction accuracy of testing data. The classification accuracy of each algorithm is calculated based on the classification results of the testing data as follows:

\begin{equation}
    \textrm{accuracy} =  \frac{\textrm{number of correct predictions}}{\textrm{number of testing data}}
\end{equation}

%METHODOLOGY
%----------------------------------------------------
\section{Methodology}
\label{methodology}
The data used in this experiment is generated using the IEEE 14-bus, IEEE 57-bus, and IEEE 118-bus systems and MATPOWER library \cite{matpower}. The measurement data consists of power flow of branches and buses which are mapped into the state variables, the voltage bus angles, using the Jacobian matrix. Based on the aforementioned process in section \ref{ssec:sys_FDI}, 10,000 instances of measurements are generated as training data with half of them being infected with an FDI attack. Another 1,000 instances are generated as testing data to calculate the classification accuracy of each method.

The experimental process consisted of two main steps. First, the classification algorithms in section \ref{ssec:ClassificationAlgorithms} are cross-validated for accuracy along varying parameters with all original features of each system. The goal is to obtain optimal parameters for each algorithm to be used for the remainder of the experiment. The second step is testing the three FS techniques described in section \ref{ssec:SL_featSelect} with the three classification algorithms described in \ref{ssec:ClassificationAlgorithms} using the optimal parameters obtained in the first step.

%EXPERIMENTAL RESULTS
%----------------------------------------------------
\section{Experimental Results}
\label{ExperimentalResults}
\begin{figure}[htbp]
\minipage{0.24\textwidth}
  \includegraphics[width=\linewidth]{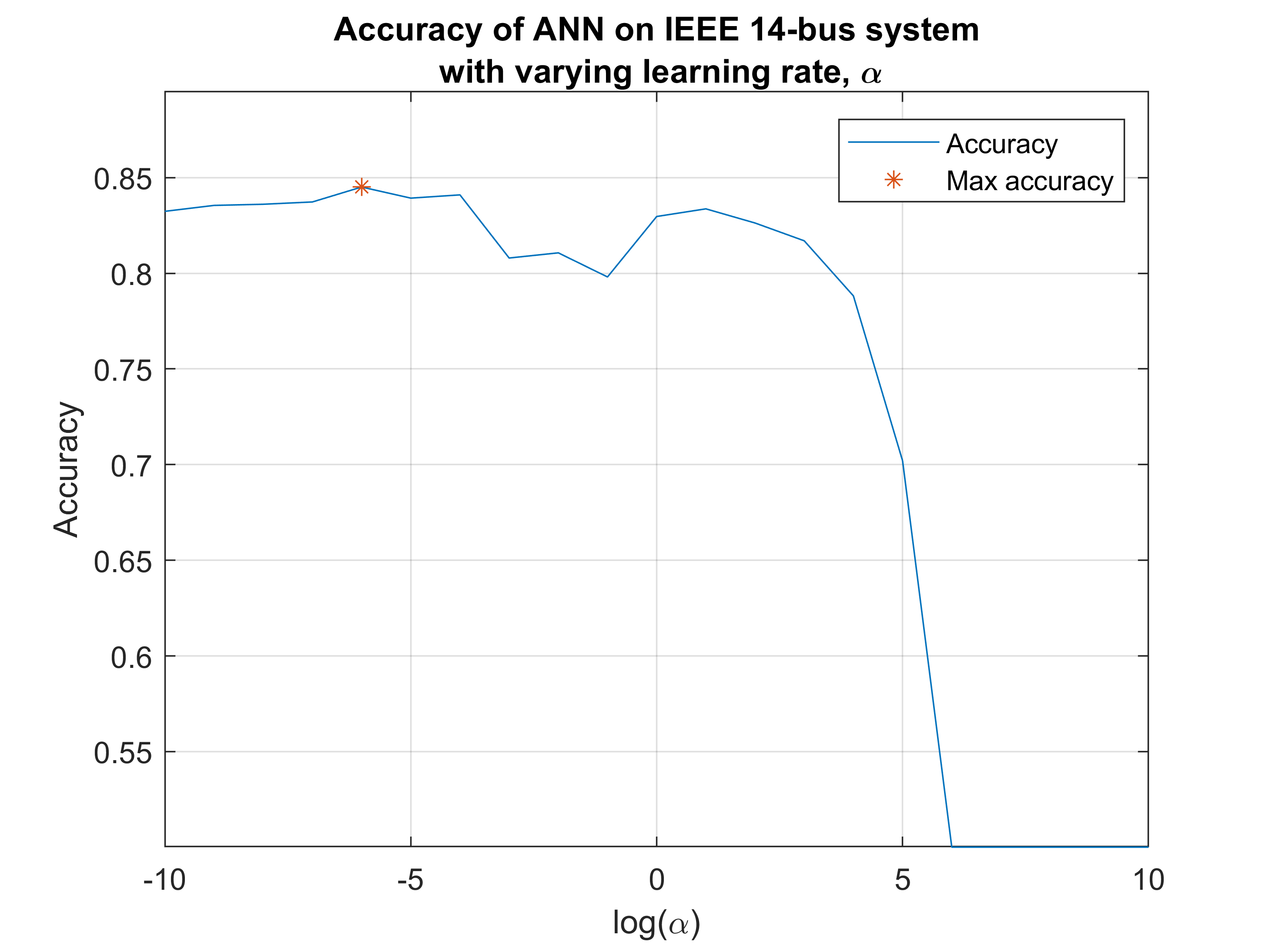}
\endminipage\hfill
\minipage{0.24\textwidth}
  \includegraphics[width=\linewidth]{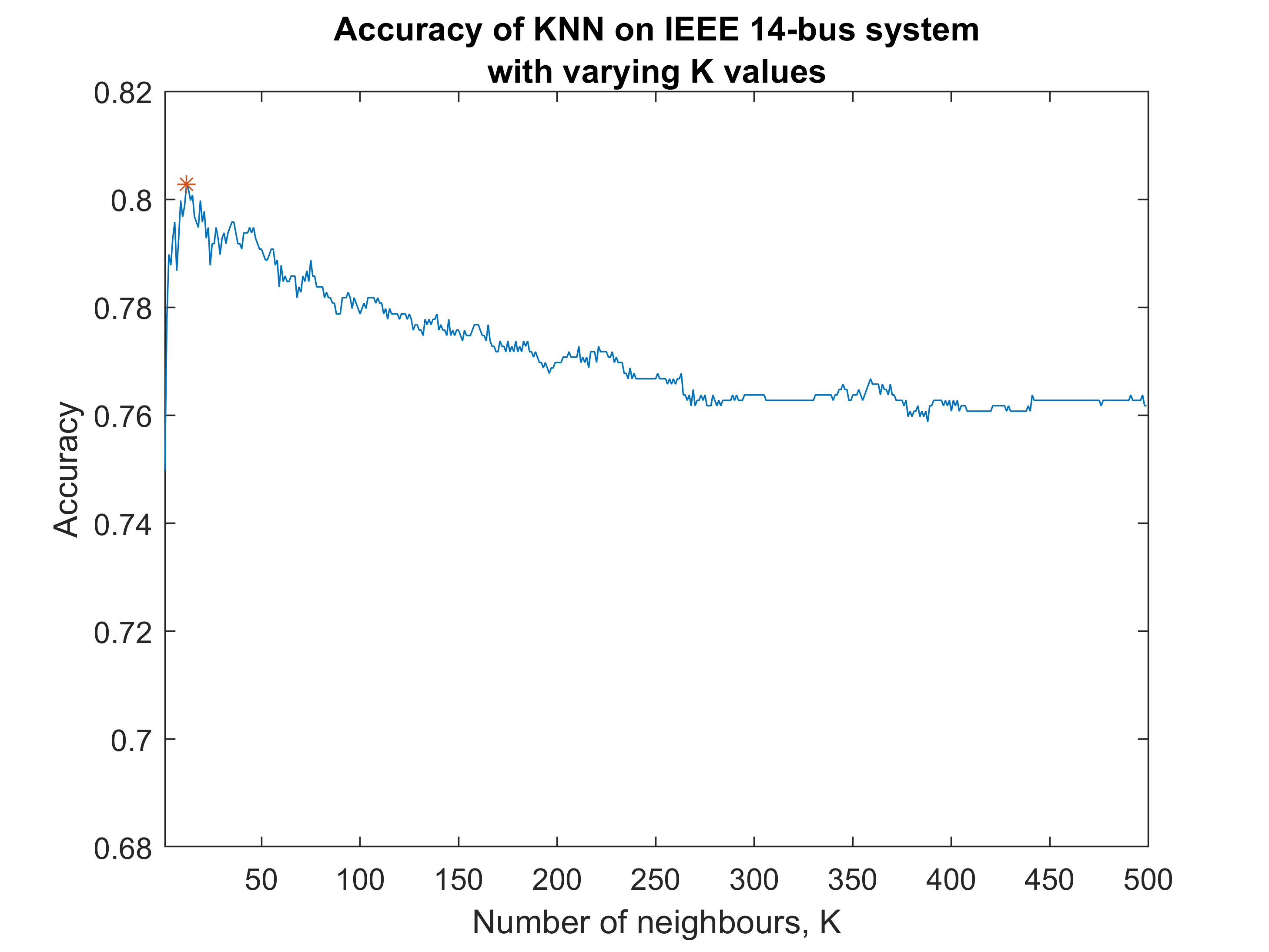}
\endminipage\hfill
\minipage{0.48\textwidth}
  \includegraphics[width=0.5\linewidth]{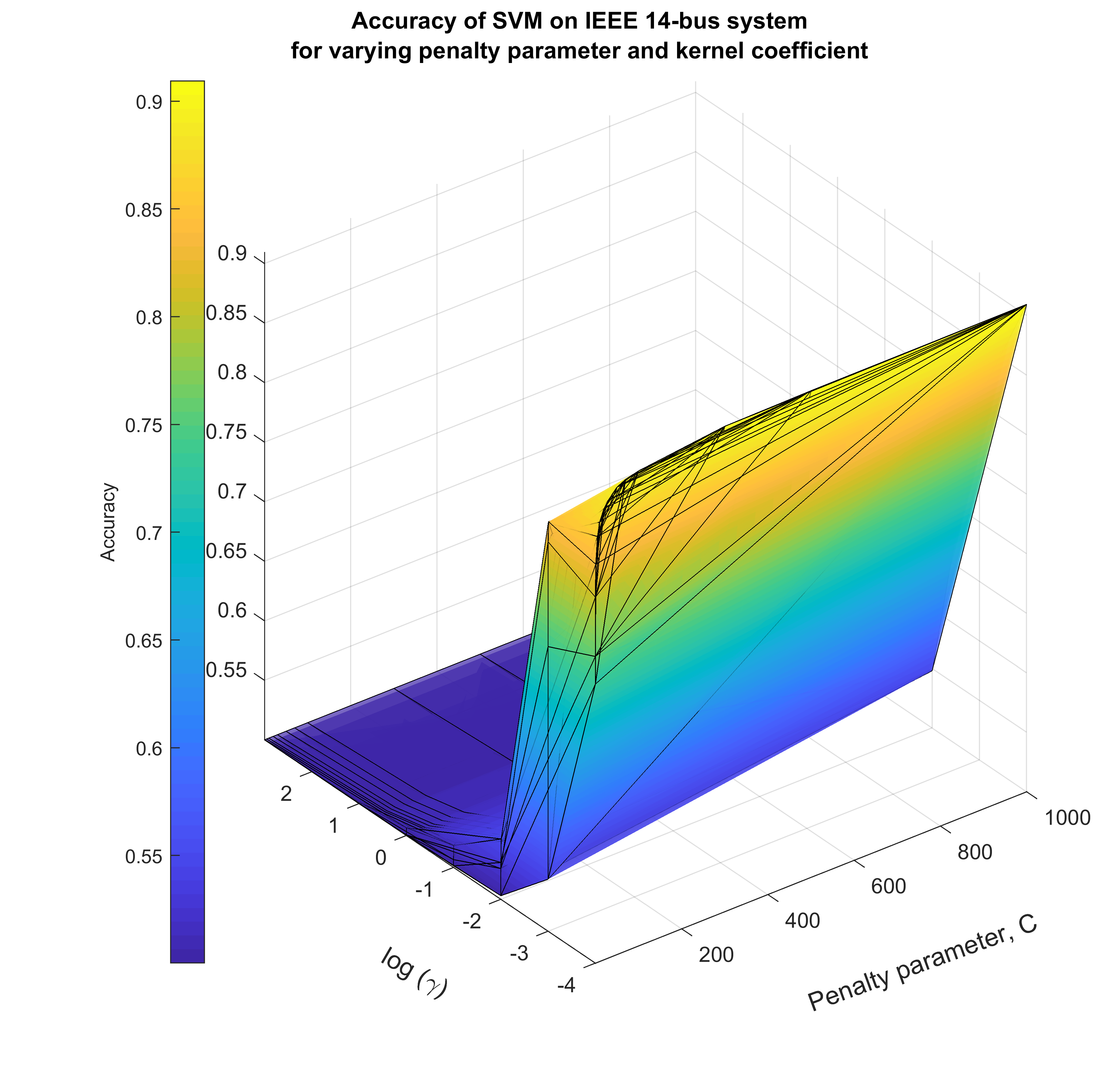}
  \centering
\endminipage
\caption{Accuracy of SVM, ANN, and KNN for varying parameters for IEEE 14-bus system.}
\label{fig:parOptResults}
\end{figure}

Parameter optimization of each of the supervised learning algorithm is performed through cross-validation of varying parameters with optimal accuracy. Figure \ref{fig:parOptResults} shows the accuracy of each of the three algorithms with varying parameters on the IEEE 14-bus system. SVM is cross-validated for varying kernel coefficient and penalty parameter,$\gamma$ and $C$ respectively, KNN is cross-validated for varying number of neighbours, $K$, and ANN is cross-validated for varying learning rate, $\alpha$. The data used for this cross-validation consists of all the measurements of the system. Optimal parameters of each learning algorithms are selected based on the maximum accuracy achieved on the IEEE 14-bus system with no FS. These parameters are stated in table \ref{tab:MLparameters}. 

%ML parameters table
\begin{table}[htbp]
\caption{Optimal parameters of the supervised learning algorithms and their corresponding accuracy on the IEEE 14-bus system with no feature selection}
\begin{center}
\begin{tabular}{|c|c|c|}
\hline
Algorithm & Parameters & Accuracy\\
\hline
SVM & $C = 1000$, $\gamma = 0.0001$ & $90.93 \%$ \\
KNN & $K = 12$ & $80.82 \%$ \\
ANN & $\alpha = 10^{-6}$ & $84.50 \%$\\
\hline
\end{tabular}
\label{tab:MLparameters}
\end{center}
\end{table}

The three FS methods, BCS, BPSO, and GA, are implemented with the parameters stated in table \ref{tab:FSparameters} which are chosen based on \cite{BCS} and \cite{GAfdi}. The resultant subset of features selected by each algorithm are tested with the three classification algorithms, SVM, KNN, and ANN, and their classification accuracy on each of the three IEEE bus systems are recorded in tables \ref{tab:IEEE14}, \ref{tab:IEEE57}, and \ref{tab:IEEE118}.

%FS Parameter Table
\begin{table}[htbp]
\caption{Parameters of the heuristic FS algorithms}
\begin{center}
\begin{tabular}{|c|c|}
\hline
\textbf{Algorithm} & \textbf{Parameters}\\
\hline
BCS & $\alpha = 0.1$, $P(a) = 0.25$, $population = 30$, $iterations = 10$ \\
\hline
BPSO & $c_1=c_2=2$, $w=0.7$, $population = 30$, $iterations = 10$ \\
\hline
GA & $mutation rate = 0.018$, $population = 50$, $iterations = 30$\\
\hline
\end{tabular}
\label{tab:FSparameters}
\end{center}
\end{table}

%IEEE14 table
\begin{table}[!htbp]
\caption{Classification accuracy of each supervised learning algorithm with each heuristic feature selection technique on the IEEE 14-bus system}
\begin{center}
\begin{tabular}{|c|c|c|c|c|c|}
\hline
\textbf{FS}&\textbf{Num of}&\multicolumn{3}{|c|}{\textbf{Classification Accuracy}} \\
\cline{3-5} 
\textbf{Method} & \textbf{Features} & \textbf{\textit{SVM}}& \textbf{\textit{KNN}}& \textbf{\textit{ANN}} \\
\hline
NO FS & $34$ & $90.79 \%$ & $80.28 \%$ & $81.78 \%$ \\
\hline
BCS & $11$ & $90.69 \%$ & $81.38 \%$ & $77.08 \%$ \\
\hline
BPSO & $8$ & $90.19 \%$ & $81.68 \%$ & $79.18 \%$ \\
\hline
GA & $8$ & $90.49 \%$ & $82.28 \%$ & $79.28 \%$ \\
\hline
\end{tabular}
\label{tab:IEEE14}
\end{center}
\end{table}

%IEEE57 table
\begin{table}[!htbp]
\caption{Classification accuracy of each supervised learning algorithm with each heuristic feature selection technique on the IEEE 57-bus system}
\begin{center}
\begin{tabular}{|c|c|c|c|c|c|}
\hline
\textbf{FS}&\textbf{Num of}&\multicolumn{3}{|c|}{\textbf{Classification Accuracy}} \\
\cline{3-5} 
\textbf{Method} & \textbf{Features} & \textbf{\textit{SVM}}& \textbf{\textit{KNN}}& \textbf{\textit{ANN}} \\
\hline
NO FS & $137$ & $88.29 \%$ & $83.08 \%$ & $50.05 \%$ \\
\hline
BCS & $94$ & $88.59 \%$ & $84.48 \%$ & $50.15 \%$ \\
\hline
BPSO & $130$ & $87.39 \%$ & $83.58 \%$ & $48.25 \%$ \\
\hline
GA & $56$ & $87.39 \%$ & $85.59 \%$ & $50.95 \%$ \\
\hline
\end{tabular}
\label{tab:IEEE57}
\end{center}
\end{table}

%IEEE118 table
\begin{table}[!htbp]
\caption{Classification accuracy of each supervised learning algorithm with each heuristic feature selection technique on the IEEE 118-bus system}
\begin{center}
\begin{tabular}{|c|c|c|c|c|c|}
\hline
\textbf{FS}&\textbf{Num of}&\multicolumn{3}{|c|}{\textbf{Classification Accuracy}} \\
\cline{3-5} 
\textbf{Method} & \textbf{Features} & \textbf{\textit{SVM}}& \textbf{\textit{KNN}}& \textbf{\textit{ANN}} \\
\hline
NO FS & $304$ & $84.88 \%$ & $74.57 \%$ & $53.05 \%$ \\
\hline
BCS & $199$ & $83.58 \%$ & $75.48 \%$ & $51.25 \%$ \\
\hline
BPSO & $160$ & $83.28 \%$ & $76.68 \%$ & $51.95 \%$ \\
\hline
GA & $122$ & $90.59 \%$ & $78.18 \%$ & $50.05 \%$ \\
\hline
\end{tabular}
\label{tab:IEEE118}
\end{center}
\end{table}

Results show that SVM and KNN are successful at detecting FDI attacks in all three IEEE bus systems. SVM is the most versatile scoring the highest classification accuracy among all the FS methods and in all three test systems. Furthermore, all three heuristic FS methods proved successful at reducing the number of features. GA produced the most successful results among the three FS methods by achieving the highest classification accuracy with minimal number of features. ANNs with the proposed architecture were unsuccessful at detecting FDI attacks regardless of the FS method.

%CONCLUSION
%----------------------------------------------------
\section{Conclusion}
\label{conclusion}
The inability of the current defence mechanisms to detect FDI attacks calls for alternative methods of detection. In this paper, supervised learning algorithms are implemented and proved to be successful at detecting FDI attacks when tested on the IEEE 14-bus, 57-bus, and 118-bus systems. Furthermore, heuristic FS methods were successful at maintaining, and sometimes increasing, the classification accuracy with significantly lower number of features. SVM and KNN algorithms proved more accurate and versatile among the three systems when compared to the ANN implemented in this paper. However, ANNs with more complex architectures are expected to have better performance on larger systems at a higher computational cost.

FS methods were all successful at increasing accuracy or reducing the number of features, and in some cases both. Classification results conclude that GA is the most efficient heuristic FS method for power systems in terms of accuracy and number of features. SVM with GA proved to be the most accurate and versatile among the three systems.

%BIBLIOGRAPHY
%----------------------------------------------------
\newpage
\bibliography{myBib.bib}

\begin{thebibliography}{10}
\providecommand{\url}[1]{#1}
\csname url@samestyle\endcsname
\providecommand{\newblock}{\relax}
\providecommand{\bibinfo}[2]{#2}
\providecommand{\BIBentrySTDinterwordspacing}{\spaceskip=0pt\relax}
\providecommand{\BIBentryALTinterwordstretchfactor}{4}
\providecommand{\BIBentryALTinterwordspacing}{\spaceskip=\fontdimen2\font plus
\BIBentryALTinterwordstretchfactor\fontdimen3\font minus
  \fontdimen4\font\relax}
\providecommand{\BIBforeignlanguage}[2]{{%
\expandafter\ifx\csname l@#1\endcsname\relax
\typeout{** WARNING: IEEEtran.bst: No hyphenation pattern has been}%
\typeout{** loaded for the language `#1'. Using the pattern for}%
\typeout{** the default language instead.}%
\else
\language=\csname l@#1\endcsname
\fi
#2}}
\providecommand{\BIBdecl}{\relax}
\BIBdecl

\bibitem{100}
H.~{Karimipour} and V.~{Dinavahi}, ``Robust massively parallel dynamic state
  estimation of power systems against cyber-attack,'' \emph{IEEE Access},
  vol.~6, pp. 2984--2995, 2018.

\bibitem{200}
H.~Karimipour and V.~Dinavahi, ``Parallel domain-decomposition-based
  distributed state estimation for large-scale power systems,'' \emph{IEEE
  Transactions on Industry Applications}, vol.~52, no.~2, pp. 1265--1269, March
  2016.

\bibitem{300}
H.~Karimipour and V.~Dinavahi, ``On false data injection attack against dynamic
  state estimation on smart power grids,'' in \emph{2017 IEEE International
  Conference on Smart Energy Grid Engineering (SEGE)}, Aug 2017, pp. 388--393.

\bibitem{FDImethods}
P.~{Chen}, S.~{Yang}, J.~A. {McCann}, J.~{Lin}, and X.~{Yang}, ``Detection of
  false data injection attacks in smart-grid systems,'' \emph{IEEE
  Communications Magazine}, vol.~53, no.~2, pp. 206--213, Feb 2015.

\bibitem{SVManomaly}
M.~{Esmalifalak}, , R.~{Zheng}, and Z.~{Han}, ``Detecting stealthy false data
  injection using machine learning in smart grid,'' in \emph{2013 IEEE Global
  Communications Conference (GLOBECOM)}, Dec 2013, pp. 808--813.

\bibitem{400}
S.~Mohammadi, V.~Desai, and H.~Karimipour, ``Multivariate mutual
  information-based feature selection for cyber intrusion detection,'' 10 2018,
  pp. 1--6.

\bibitem{OzayML}
M.~{Ozay}, I.~{Esnaola}, F.~T. {Yarman Vural}, S.~R. {Kulkarni}, and H.~V.
  {Poor}, ``Machine learning methods for attack detection in the smart grid,''
  \emph{IEEE Transactions on Neural Networks and Learning Systems}, vol.~27,
  no.~8, pp. 1773--1786, Aug 2016.

\bibitem{SVMandKNN}
J.~{Yan}, B.~{Tang}, and H.~{He}, ``Detection of false data attacks in smart
  grid with supervised learning,'' in \emph{2016 International Joint Conference
  on Neural Networks (IJCNN)}, July 2016, pp. 1395--1402.

\bibitem{StateEstimation}
M.~Ozay, I.~Esnaola, F.~T. Yarman-Vural, S.~R. Kulkarni, and H.~V. Poor,
  ``Sparse attack construction and state estimation in the smart grid:
  Centralized and distributed models,'' \emph{IEEE Journal on Selected Areas in
  Communications}, vol.~31, pp. 1306--1318, 2013.

\bibitem{500}
H.~{Karimipour} and V.~{Dinavahi}, ``Extended kalman filter-based parallel
  dynamic state estimation,'' \emph{IEEE Transactions on Smart Grid}, vol.~6,
  no.~3, pp. 1539--1549, May 2015.

\bibitem{600}
H.~{Karimipour} and V.~{Dinavahi}, ``Parallel relaxation-based joint dynamic
  state estimation of large-scale power systems,'' \emph{IET Generation,
  Transmission Distribution}, vol.~10, no.~2, pp. 452--459, 2016.

\bibitem{FDI}
R.~Bobba, K.~Davis, Q.~Wang, H.~Khurana, K.~Nahrstedt, and T.~J~Overbye,
  ``Detecting false data injection attacks on dc state estimation,'' 01 2010.

\bibitem{700}
S.~Mohammadi, H.~Mirvaziri, M.~Ghazizadeh-Ahsaee, and H.~Karimipour, ``Cyber
  intrusion detection by combined feature selection algorithm,'' \emph{Journal
  of Information Security and Applications}, vol.~44, pp. 80--88, 02 2019.

\bibitem{CSoriginal}
X.~Y. and, ``Cuckoo search via lévy flights,'' in \emph{2009 World Congress on
  Nature Biologically Inspired Computing (NaBIC)}, Dec 2009, pp. 210--214.

\bibitem{BCS}
D.~{Rodrigues}, L.~A.~M. {Pereira}, T.~N.~S. {Almeida}, J.~P. {Papa}, A.~N.
  {Souza}, C.~C.~O. {Ramos}, and X.~{Yang}, ``Bcs: A binary cuckoo search
  algorithm for feature selection,'' in \emph{2013 IEEE International Symposium
  on Circuits and Systems (ISCAS2013)}, May 2013, pp. 465--468.

\bibitem{GAfdi}
S.~Ahmed, Y.~Lee, S.~Hyun, and I.~Koo, ``Covert cyber assault detection in
  smart grid networks utilizing feature selection and euclidean distance-based
  machine learning,'' \emph{Applied Sciences}, vol.~8, p. 772, 05 2018.

\bibitem{PSOclassification}
B.~{Xue}, M.~{Zhang}, and W.~N. {Browne}, ``Particle swarm optimization for
  feature selection in classification: A multi-objective approach,'' \emph{IEEE
  Transactions on Cybernetics}, vol.~43, no.~6, pp. 1656--1671, Dec 2013.

\bibitem{SVM}
\BIBentryALTinterwordspacing
N.~Guenther and M.~Schonlau, ``Support vector machines,'' \emph{The Stata
  Journal}, vol.~16, no.~4, pp. 917--937, 2016. [Online]. Available:
  \url{https://doi.org/10.1177/1536867X1601600407}
\BIBentrySTDinterwordspacing

\bibitem{matpower}
R.~D. {Zimmerman}, C.~E. {Murillo-Sanchez}, and R.~J. {Thomas}, ``Matpower:
  Steady-state operations, planning, and analysis tools for power systems
  research and education,'' \emph{IEEE Transactions on Power Systems}, vol.~26,
  no.~1, pp. 12--19, Feb 2011.

\end{thebibliography}
\bibliographystyle{IEEEtran}
\end{document}